\documentclass[prl, twocolumn, superscriptaddress,floatfix, nofootinbib, preprintnumbers]{revtex4-2}
\usepackage{dcolumn}
\usepackage{bm}
\usepackage{color}
\usepackage{amssymb}
\usepackage{amsmath}
\usepackage{graphicx}
\usepackage{amsfonts}
\usepackage{slashed}
\usepackage{pstricks}
\usepackage{float}
\usepackage[colorlinks = true,
            linkcolor = blue,
            urlcolor  = blue,
            citecolor = blue,
            anchorcolor = blue]{hyperref}
\usepackage{array}
\allowdisplaybreaks
\usepackage{dcolumn}
\usepackage{epsf}
\usepackage{slashed}
\usepackage{tikz}
\usepackage[compat=1.1.0]{tikz-feynman}
\usepackage{feynmf}
\graphicspath{{./figs/}}

\usepackage{color}
\definecolor{gray}{rgb}{0.6,0.6,0.6}
\definecolor{darkgreen}{rgb}{0.0, 0.545098, 0.0}
\definecolor{darkblue}{rgb}{0.0, 0.0, 0.545098}

\usepackage{ulem}

\begin{document}
\title{\textsc{ResBos2} and the CDF W Mass Measurement}
\author{{Joshua} Isaacson}
\email{isaacson@fnal.gov}
\affiliation{
Theoretical Physics Department, Fermi National Accelerator Laboratory, P.O. Box 500, Batavia, IL 60510, USA
}
\author{Yao Fu}
\affiliation{Department of Modern Physics, University of Science and Technology of China, Jinzhai Road 96, Hefei, Anhui, 230026, China}
\author{C.-P. Yuan}
\affiliation{Department of Physics and Astronomy, Michigan State University, 567 Wilson Road, East Lansing, MI 48824, USA}
\preprint{FERMILAB-PUB-22-374-T, MSUHEP-22-017}

\begin{abstract} 
The recent CDF $W$ mass measurement of 80,433 $\pm$ 9 MeV is the most precise direct measurement. However, this result deviates from the Standard Model predicted mass of 80,359.1 $\pm$ 5.2 MeV by $7\sigma$. The CDF experiment used an older version of the \textsc{ResBos} code that was only accurate at NNLL+NLO, while the \textsc{ResBos2} code is able to make predictions at N${}^3$LL+NNLO accuracy. We determine that the data-driven techniques used by CDF capture most of the higher order corrections, and using higher order corrections would result in a decrease in the value reported by CDF by at most 10 MeV.
\end{abstract}
\maketitle

In the Standard Model (SM) of particle physics, the electroweak sector can be uniquely determined given 3 input parameters to fix the values of two gauge couplings and the vacuum expectation value, the Higgs and fermion masses, and the weak mixing matrices.
One of the predicted results is the mass of the $W$ boson. Currently, the electroweak global fits predict the mass to be 80,359.1 $\pm$ 5.2 MeV~\cite{deBlas:2021wap}.
Recently, the CDF experiment reported the most precise direct measurement of the $W$ mass as: 80,433 $\pm$ 9 MeV~\cite{CDF:2022hxs}. 
This corresponds to a $7\sigma$ deviation from the SM, and has spurred many Beyond the Standard Model explanations~\cite{Fan:2022dck,Lu:2022bgw,Athron:2022qpo,Yang:2022gvz,Tang:2022pxh,Du:2022pbp,Cacciapaglia:2022xih,Blennow:2022yfm,Sakurai:2022hwh,Arias-Aragon:2022ats,Liu:2022jdq,Babu:2022pdn,DiLuzio:2022xns,Heckman:2022the,Lee:2022nqz,Bahl:2022xzi,Song:2022xts,Athron:2022isz,Heo:2022dey,Crivellin:2022fdf,Endo:2022kiw,Du:2022brr,Cheung:2022zsb,DiLuzio:2022ziu,Balkin:2022glu,Biekotter:2022abc,Han:2022juu,Zheng:2022irz,Ahn:2022xeq,Kawamura:2022uft,Ghoshal:2022vzo,FileviezPerez:2022lxp,Nagao:2022oin,Kanemura:2022ahw,Zhang:2022nnh,Borah:2022obi,Chowdhury:2022moc,Arcadi:2022dmt,Popov:2022ldh,Carpenter:2022oyg,Ghorbani:2022vtv,Bhaskar:2022vgk,Du:2022fqv,Borah:2022zim,Cao:2022mif,Heeck:2022fvl,Lee:2022gyf,Batra:2022pej,Faraggi:2022emm,Abouabid:2022lpg,Basiouris:2022wei,Wang:2022dte,Dcruz:2022dao}.
However, the ATLAS and LHCb experiments also have directly measured the mass of the $W$ boson and found results of 80,370 $\pm$ 19 MeV~\cite{ATLAS:2017rzl} and 80,354 $\pm$ 32 MeV~\cite{LHCb:2021bjt}, respectively. 
The tension between the direct measurements are at a level of $3\sigma$, and have raised concerns about the CDF measurement.
One major concern brought up is that an older version of the \textsc{ResBos} code~\cite{Balazs:1997xd,Landry:2002ix} was used, which is only at the accuracy of next-to-next-to-leading logarithimic accuracy matched to a next-to-leading fixed order calculation (NNLL+NLO)~\cite{Castelvecchi2022-zl}.
Here we investigate the shift in the CDF result that would occur if the N${}^3$LL+NNLO accurate \textsc{ResBos} version 2.0~\cite{Isaacson:2017hgb} calculation was used instead, including the correct angular functions at NNLO in Quantum Chromodyamics. We will denote \textsc{ResBos} version 2.0 as \textsc{ResBos2} for the remainder of the text. Additionally, we consider the effects on the PDF uncertainty from using the higher order prediction. State-of-the-art calculations are now available in other codes at N${}^3$LL'+N${}^3$LO accuracy~\cite{Chen:2022cgv,Bizon:2018foh,Bizon:2019zgf,Re:2021con}.

The \textsc{ResBos2} calculation includes the cusp anomalous dimension at $\mathcal{O}(\alpha_s^4)$, the non-cusp anomalous dimension at $\mathcal{O}(\alpha_s^3)$, the hard collinear coefficient at $\mathcal{O}(\alpha_s^2)$, and is matched to the fixed order calculation at NNLO. This new accuracy corresponds to an accuracy of N${}^3$LL+NNLO as described in Tab.~\ref{tab:ResummationOrder}. The CDF experiment chose to use the same version of \textsc{ResBos} as was used in their previous analysis~\cite{CDF:2007mxw} when the higher order corrections were unknown. The \textsc{ResBos} codebases implement the Collins-Soper-Sterman (CSS) resummation formalism~\cite{Collins:1984kg,Collins:2011zzd}, which performs the transverse momentum resummation in $b$-space as
\begin{align}\label{eq:resum}
    \frac{{\rm d}\sigma}{{\rm d}Q^2\,{\rm d}^2\vec{p}_T\,{\rm d}y\,{\rm d}\cos\theta\,{\rm d}\phi} &= \sigma_0 \int \frac{{\rm d}^2 b}{(2\pi)^2} e^{i \vec{p}_T \cdot \vec{b}} e^{-S(b)} \\
    & \times C \otimes f(x_1, \mu)\, C \otimes f(x_2, \mu) \nonumber \\
    & + Y(Q, \vec{p}_T, x_1, x_2, \mu_R, \mu_F) \nonumber\, ,
\end{align}
where $Q$, $p_T$ and $y$ are respectively the invariant mass, transverse momentum and rapidity of the lepton system, $\theta (\phi)$ is the polar (azimuthal) angle in the Collins-Soper frame~\cite{Collins:1977iv}, $\sigma_0$ is the leading order matrix element,
$x_{1,2} = \sqrt{(Q^2+p_T^2)/s}\,e^{\pm y}$, with $s$ being the center-of-mass energy of the collider and $Y$ contains the finite terms of the fixed order calculation in the limit $p_T \rightarrow 0$. The Sudakov factor ($S(b)$) is defined as
\begin{equation}\label{eq:sud}
    S(b) = \int^{C_2^2 Q^2}_{C_1^2/b^2} \frac{{\rm d}\bar{\mu}^2}{\bar{\mu}^2} \left[\ln\left(\frac{C_2^2 Q^2}{\bar{\mu}^2}\right) A\left(\bar{\mu}, C_1\right) + B\left(\bar{\mu}, C_1, C_2\right)\right]\, ,
\end{equation}
and $C\otimes f(x,\mu=C_3/b)$ represents the convolution of the hard collinear kernel with the PDF, the values of $A$, $B$, and $C$ can be calculated order-by-order in perturbation theory. Additional details and the values of the coefficients needed for N${}^3$LL accuracy are given in Appendix~\ref{app:resum}.

Another area of concern is the handling of the angular coefficients within the \textsc{ResBos} code. The angular coefficients are given by
\begin{widetext}
\begin{align}
    \frac{{\rm d}\sigma}{{\rm d}p_T\,{\rm d}y\,{\rm d}Q\,{\rm d}\cos\theta\,{\rm d}\phi} &= \frac{3}{16\pi}\frac{{\rm d}\sigma}{{\rm d}p_T\,{\rm d}y\,{\rm d}Q}  \nonumber \\
    & \times\left\{\left(1+\cos^2\theta\right) + \frac{1}{2}A_0\left(1-3\cos^2\theta\right)+A_1\sin 2\theta \cos\phi + A_2\sin^2\theta \cos 2\phi \right. \\
    & \left. +A_3\sin\theta\cos\phi + A_4 \cos\theta + A_5 \sin^2\theta\sin 2\phi + A_6 \sin 2\theta\sin\phi + A_7 \sin\theta \sin\phi\right\}\, . \nonumber
\end{align}
\end{widetext}
The coefficients $A_i$ are determined order-by-order in the fixed order calculation, with only $A_4$ being non-zero at leading order. The coefficients $A_{5,6,7}$ are zero until NNLO and are negligible afterwards, and will therefore be ignored in this study. 
While the matching to NLO was exact in terms of the angular coefficients, it was pointed out that the original matching to NNLO in the \textsc{ResBos} code did not correctly reproduce the breaking of the Lam-Tung relation~\cite{Lam:1980uc}.
To account for this, the \textsc{ResBos2} code uses $k$-factors obtained from \textsc{MCFM}~\cite{Campbell:2015qma} to correctly reproduce the angular distributions at NNLO accuracy. This effect is expected to be a small effect in the extraction of $M_W$ since the CDF experiment applies a cut of $p_T(W) < 15$ GeV. (See App.~\ref{app:angular}). In the small $p_T$ region all angular coefficients tend towards zero with the exception of the overall leading coefficient and $A_4$.

\begin{table*}[ht]
    \centering
    \begin{tabular}{| c | c | c c | c |}
    \hline
    & & \multicolumn{2}{c |}{Anomalous Dimension} & \\
    Order & Boundary Condition (C) & $\gamma_i$ (B) & $\Gamma_{cusp}$ (A) & Fixed Order Matching (Y)  \\
    \hline
    LL & 1 & - & 1-loop & - \\
    NLL & 1 & 1-loop & 2-loop & - \\
    NLL' (+ NLO) & $\alpha_s$ & 1-loop & 2-loop & $\alpha_s$ \\
    NNLL (+ NLO) & $\alpha_s$ & 2-loop & 3-loop & $\alpha_s$ \\
    NNLL' (+ NNLO) & $\alpha_s^2$ & 2-loop & 3-loop & $\alpha_s^2$ \\
    N${}^3$LL (+ NNLO) & $\alpha_s^2$ & 3-loop & 4-loop & $\alpha_s^2$ \\
    N${}^3$LL' (+ N${}^3$LO) & $\alpha_s^3$ & 3-loop & 4-loop & $\alpha_s^3$ \\
    N${}^4$LL (+ N${}^3$LO) & $\alpha_s^3$ & 4-loop & 5-loop & $\alpha_s^3$ \\
    \hline
    \end{tabular}
    \caption{The definitions for the accuracy of the resummation calculation. The accuracy used by CDF was NNLL + NLO, while the state-of-the-art is N${}^{3}$LL + NNLO.}
    \label{tab:ResummationOrder}
\end{table*}

To estimate the effects of higher order corrections, we generate pseudoexperiments for the $Z$ and $W$ boson at N${}^{3}$LL+NNLO accuracy with correct angular distributions.
The pseudoexperiments use the \texttt{CT18NNLO} pdf~\cite{Hou:2019efy} and fix the non-perturbative function to the BLNY global fit values~\cite{Landry:2002ix}. These two choices should not have
a significant impact on the outcome of the study. The $Z$ boson events are generated satisfying $p_T(Z) < 15$ GeV, $30 < p_T(\ell) < 55$ GeV, $\lvert\eta(\ell)\rvert < 1$, and $66 < M_{\ell\ell} < 116$ GeV. The $W$ boson events are generated satisfying $p_T(W) < 15$ GeV, $30 < p_T(\ell) < 55$ GeV, $30 < p_T(\nu) < 55$ GeV, $\lvert\eta(\ell)\rvert < 1$, and $60 < m_T < 100$ GeV. Here, $m_T$ is the transverse mass and is defined as
\begin{equation}\label{eq:mt}
    m_T^2 =  2 \left(p_T(\ell) p_T(\nu) - \vec{p}_T(\ell) \cdot \vec{p}_T(\nu)\right),
\end{equation}
where $\vec{p}_T(\ell,\nu)$ is the vector transverse momentum and the dot product is related to the difference in $\phi$ angle between the two vectors.
These selection criteria are consistent with those used by CDF.
After the selection criteria, CDF had a selection of 1,811,700 (66,180) $W \rightarrow e\nu$ ($Z \rightarrow ee$) events and 2,424,486 (238,534) $W \rightarrow \mu\nu$ ($Z \rightarrow \mu\mu$) events~\cite{CDF:2022hxs}. When displaying the statistical uncertainty from CDF in figures, we will use the electron numbers since the background and effects from final state radiation (FSR) are significantly smaller.
The main difference is the cut on the $p_T(Z)$, which was chosen to be 15 GeV vs. 30 GeV. This choice should not have an impact on the final extracted mass, and was made to appropriately tune the prediction with the information provided by CDF. The width of the $W$ boson was fixed to 2.0895 GeV to be consistent with the value used by CDF~\cite{CDF:2022hxs}, the impact of the width has a minor effect on the extraction of the mass (see App.~\ref{app:width}), and thus fixing the value does not impact the conclusions of this work.

\begin{figure}[ht]
    \centering
    \includegraphics[width=0.47\textwidth]{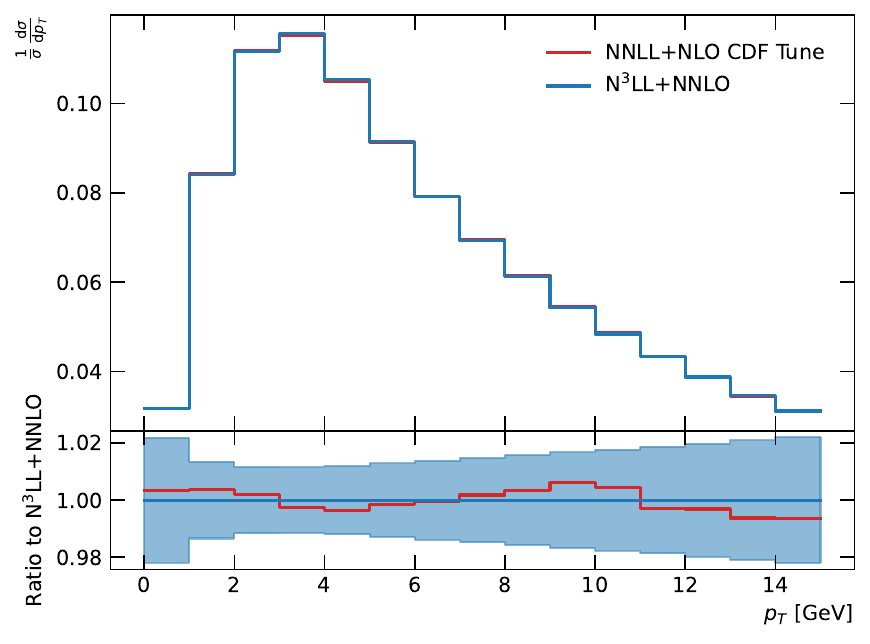}
    \caption{Comparison of the generated pseudodata for the $Z$ transverse momentum for the state-of-the-art \textsc{ResBos2} calculation at N${}^3$LL+NNLO (blue) accuracy compared to the NNLL+NLO (red) prediction tuned to the $Z$ transverse momentum distribution. The blue band represents the statistical uncertainty of the CDF result.}
    \label{fig:ZTune}
\end{figure}

The CDF experiment used data driven techniques to tune the \textsc{ResBos} prediction to reproduce the $p_T(Z)$ data.
The tuning of the \textsc{ResBos} prediction involved modifying the values of the $g_2$ parameter in the BLNY functional form and adjusting the value of $\alpha_s(M_Z)$ used.
To mimic the procedure carried out by CDF, we fit the value of $g_2$ and $\alpha_s(M_Z)$ in the NNLL+NLO prediction to reproduce the transverse momentum spectrum of the $Z$ boson at N${}^3$LL+NNLO, and validate it against the $W$ boson transverse momentum. We find that the best fit result comes from using $g_2 = 0.662$ GeV$^{2}$ and $\alpha_s(M_Z) = 0.120$, as shown in Fig.~\ref{fig:ZTune}. To correctly account for the modified value of $\alpha_s(M_Z)$, we used the \texttt{CT18NNLO\_as\_0120} PDF set for the \textsc{ResBos} prediction at NNLL+NLO.

\begin{figure}
    \centering
    \includegraphics[width=0.47\textwidth]{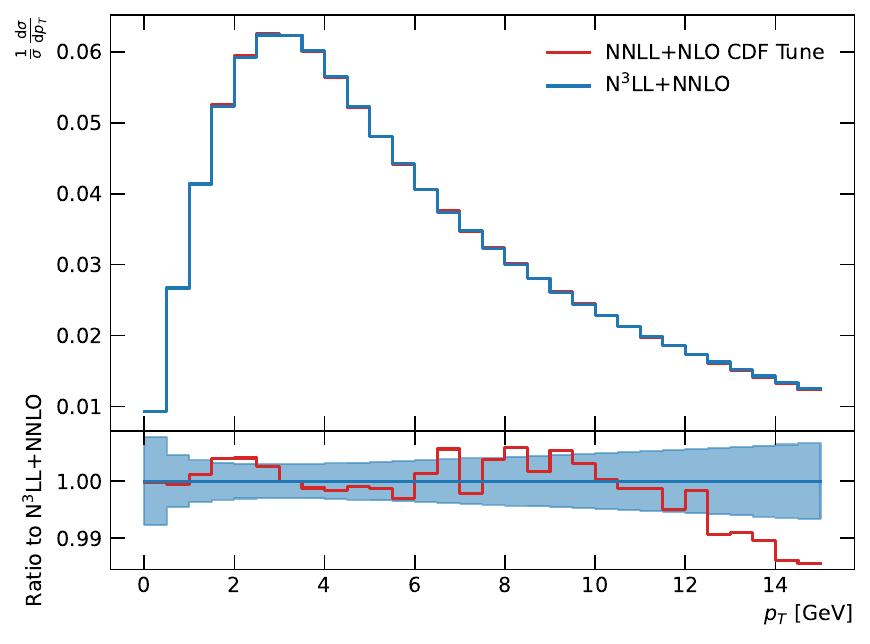}
    \caption{Comparison of the generated pseudodata for the $W$ transverse momentum for the state-of-the-art \textsc{ResBos2} calculation at N${}^3$LL+NNLO (blue) accuracy compared to the NNLL+NLO (red) prediction tuned to the $Z$ transverse momentum distribution. The blue band represents the statistical uncertainty of the CDF result.}
    \label{fig:WTune}
\end{figure}
\begin{figure}[ht]
    \centering
    \includegraphics[width=0.47\textwidth]{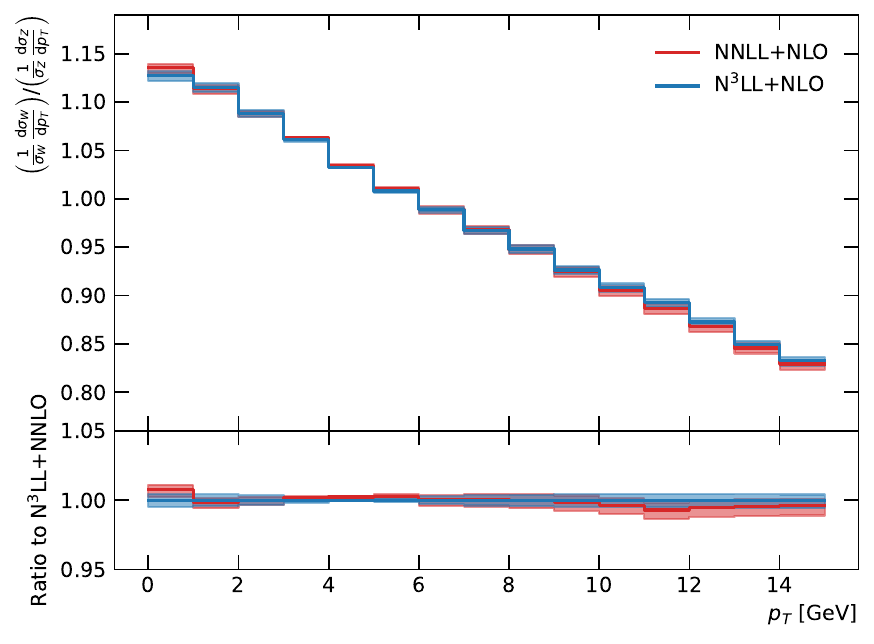}
    \caption{The ratio for the normalized $W$ transverse momentum to the normalized $Z$ transverse momentum in the region used by the CDF experiment. The NNLL+NLO (red) prediction and the N${}^3$LL+NNLO (blue) prediction are consistent with each other over this region. The scale uncertainty is treated in a fully correlated manner with the 15 scale scheme.
}
    \label{fig:PTratio}
\end{figure}

Another major concern related to the prediction of the $p_T(Z)$ and $p_T(W)$ distribution is in the effect higher order corrections have on their ratio.
We investigate the shift in this ratio in Fig.~\ref{fig:PTratio}. We take the more aggressive estimate of the uncertainty using the correlated 15 scale variation scheme (See App.~\ref{app:resum} for more details on the scales). With this aggressive estimate, we find that the ratio is not very sensitive to higher order corrections.
Therefore, we expect the impact of using the state-of-the-art \textsc{ResBos2} prediction for this ratio will only result in a negligible effect on the extracted $W$ mass. The CDF experiment~\cite{CDF:2022hxs} used \textsc{DYQT}~\cite{Bozzi:2008bb,Bozzi:2010xn} to estimate the uncertainty induced by this ratio (See App.~\ref{app:ratio_pT} for a similar ResBos analysis).

\begin{figure}[ht]
    \includegraphics[width=0.47\textwidth]{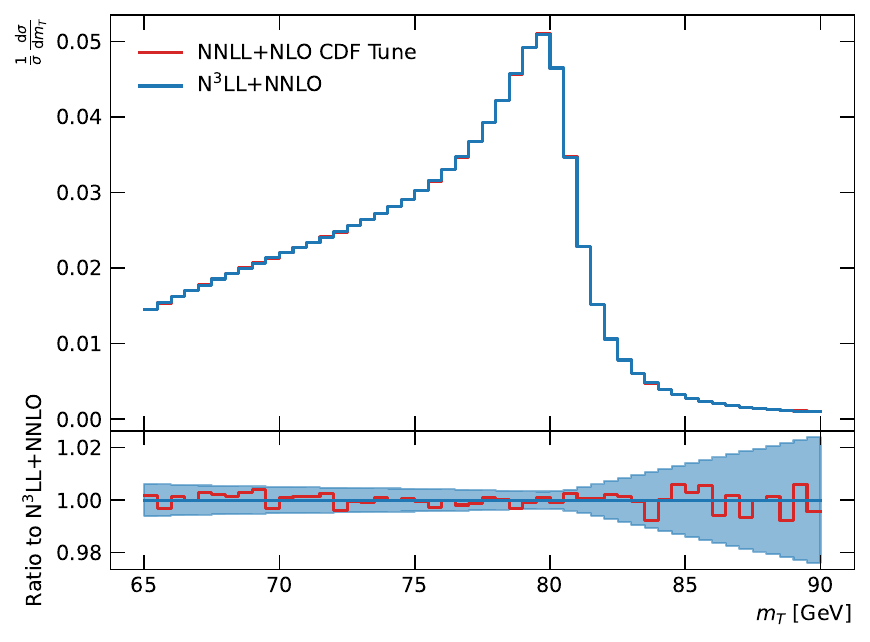}
    \caption{$W$ mass fit results to the pseudoexperiment for $m_T$. The pseudodata is generated at N${}^3$LL+NNLO accuracy with the default BLNY parametrization. The tuned NNLL+NLO results are then used for a template fit to extract the $W$ mass~\cite{CDF:2022hxs}. The tuning resulted in a best fit value of $g_2 = 0.66$ GeV${}^{-2}$ and $\alpha_s(M_Z)=0.120$. The best fit mass (80,386 MeV) is shown in red. The blue band represents the statistical uncertainty of the CDF result. Detector effects and FSR are not included here, but the corresponding result for $m_T$ can be found in Appendix~\ref{app:detector}.}
    \label{fig:Wboson}
\end{figure}

\begin{figure}[htbp]
    \includegraphics[width=0.47\textwidth]{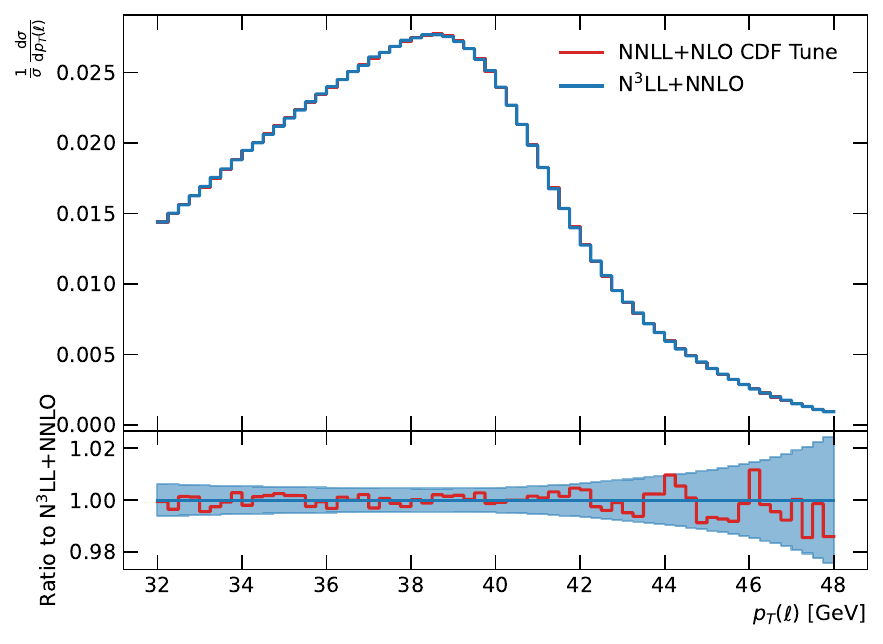}
    \caption{Similar to Fig.~\ref{fig:Wboson} but using $p_T(\ell)$ to extract the $W$ mass. The best fit mass (red) is 80,388 MeV.}
    \label{fig:Wboson2}
\end{figure}

\begin{figure}[htbp]
    \includegraphics[width=0.47\textwidth]{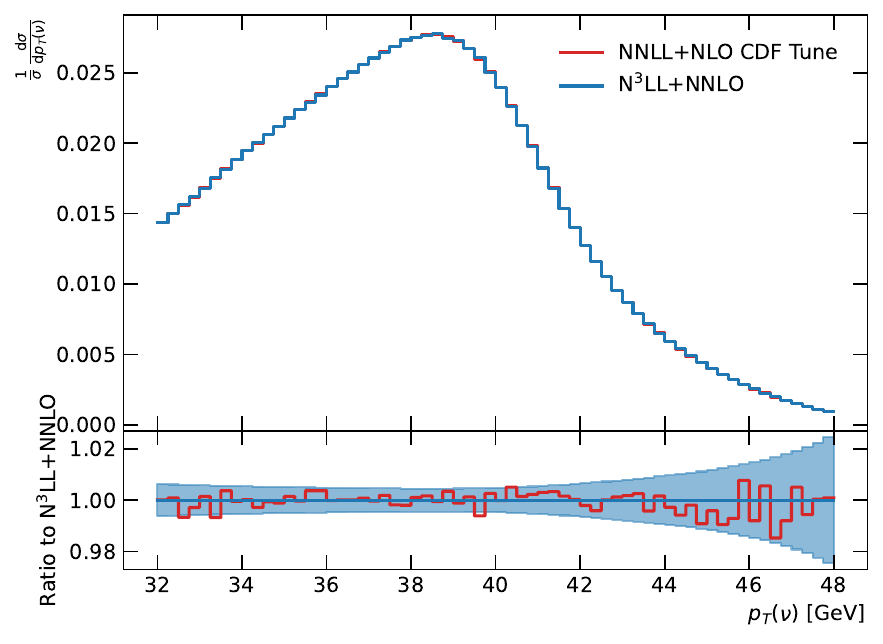}
    \caption{Similar to Fig.~\ref{fig:Wboson} but using $p_T(\nu)$ to extract the $W$ mass. The best fit mass (red) is 80,389 MeV.}
    \label{fig:Wboson3}
\end{figure}

The tuned prediction is then used to produce a series of templates to be used to extract the $W$ boson mass in $m_T$, $p_T(\ell)$, and $p_T(\nu)$, this is based on the procedure used by CDF~\cite{CDF:2022hxs}. We find that the higher order corrections can induce a shift in the $W$ boson mass of less than 10 MeV, with the result being consistent with zero within the statistical accuracy of CDF. The results are given for each observable in Tab.~\ref{tab:HOresults}. The shift without detector and final state QED radiation (FSR) effects is largest for the $p_T(\nu)$ distribution with a shift of 6.6 MeV, while including detector effects reduces the shift for all observables but $m_T$. Details on the detector smearing effects can be found in Appendix~\ref{app:detector}. The comparison of the tuned result to the pseudoexperiment is shown in Figs.~\ref{fig:Wboson},~\ref{fig:Wboson2}, and~\ref{fig:Wboson3}. Thus, we conclude that the tuning of the NNLL+NLO prediction to reproduce the CDF $Z$ transverse momentum data helps to capture the most important higher order corrections. We also conclude that redoing the analysis will not resolve the tension between CDF and the SM, and at most would amount in a decrease in the CDF value by 10 MeV.

\begin{table}[ht]
    \centering
    \begin{tabular}{|c|c|c|}
        \hline
        & \multicolumn{2}{c|}{Mass Shift [MeV]} \\
        \hline
        Observable & \textsc{ResBos2} & +Detector Effect+FSR \\
        \hline
        $m_T$ & 1.5 $\pm$ 0.5 & 0.2  $\pm$ 1.8 $\pm$ 1.0\\
        $p_T(\ell)$ & 3.1 $\pm$ 2.1 & 4.3 $\pm$ 2.7 $\pm$ 1.3 \\
        $p_T(\nu)$ & 4.5 $\pm$ 2.1 & 3.0  $\pm$ 3.4 $\pm$ 2.2\\
        \hline
    \end{tabular}
    \caption{Summary of the shift in $M_W$ due to higher order corrections. For reference, the CDF result was 80,433 $\pm$ 9 MeV~\cite{CDF:2022hxs} and the SM predicted value is 80,359.1 $\pm$ 5.2 MeV~\cite{deBlas:2021wap}. The second column shows the shift in the mass neglecting detector effects and final state radiation (FSR), while the third column includes an estimate for detector effects and FSR in the mass shift. The first uncertainty is the statistical uncertainty induced in the mass extraction due to the number of \textsc{ResBos} events generated for the pseudoexperiments and the mass templates. The second uncertainty is the detector effect uncertainty calculated by using 100 different smearings of the data to extract the $W$ mass. Additional details on the smearing can be found in Appendix~\ref{app:detector}.}
    \label{tab:HOresults}
\end{table}

Finally, concerns pertaining to the PDF uncertainty quoted by CDF were raised as being too small. CDF claims an uncertainty from the PDFs of 3.9 MeV~\cite{CDF:2022hxs} from the \texttt{NNPDF3.1 NNLO} set~\cite{NNPDF:2017mvq}. With a shift in the central value of $\pm 2$ MeV~\cite{CDF:2022hxs} from considering \texttt{CT18}~\cite{Hou:2019efy} and \texttt{MMHT2014}~\cite{Harland-Lang:2014zoa} at NNLO. Additionally, CDF looked at the impact from the NLO sets for \texttt{ABMP16}~\cite{Alekhin:2016uxn}, \texttt{CJ15}~\cite{Accardi:2016qay}, \texttt{MMHT2014}~\cite{Harland-Lang:2014zoa}, \texttt{NNPDF3.1}~\cite{NNPDF:2017mvq}, and \texttt{CT18}~\cite{Hou:2019efy}. Here they found a shift of $\pm 3$ MeV neglecting \texttt{CT18}, and $\pm 6$ MeV including \texttt{CT18}. The \textsc{ResBos} prediction used by CDF used the \texttt{CTEQ6M} PDF~\cite{Pumplin:2002vw}, and they found the mass shift from \texttt{CTEQ6M} to \texttt{NNPDF3.1 NNLO} was $+(3.3, 3.6, 3.0)$ MeV for the ($m_T,p_T^\ell,p_T^\nu$) fits, respectively~\cite{CDF:2022hxs}. We repeated the study of the PDF uncertainties at  N${}^3$LL+NNLO for \texttt{NNPDF3.1}, \texttt{MMHT2014}, and \texttt{CT18} at NNLO and NLO, and \texttt{CTEQ6M} at NLO using the technique described in Ref.~\cite{Hussein:2019kqx}. We find that the PDF uncertainty from \texttt{CT18NNLO} is (1.3, 15.9, 15.5) MeV without including detector effects for ($m_T$, $p_T(\ell)$, and $p_T(\nu)$). In the CDF fit, the total combination is dominated by the $m_T$ result ($\sim 65\%$), with the charged lepton $p_T$ channel accounting for $\sim 25.4\%$~\cite{CDF:2022hxs}. Since it is unclear how to appropriately propagate the uncertainties from individual observables to the final mass extraction, we just quote the results from the dominant $m_T$ observable. The shifts found in the $W$ mass using $m_T$ for the various PDFs are shown in Tab.~\ref{tab:pdf}, with additional details for the other observables in Appendix~\ref{app:pdf_correlations}. 
The PDF uncertainties that we find are consistent with the 3.9 MeV quoted by CDF~\cite{CDF:2022hxs}.

\begin{table}[htbp]
    \centering
    \begin{tabular}{|c|c|c|}
        \hline
         &\multicolumn{2}{c|}{$m_T$}  \\
        \hline
        PDF Set & NNLO & NLO   \\
        \hline
        \texttt{CT18} & 0.0 $\pm$ 1.3 & 1.8 $\pm$ 1.2 \\
        \texttt{MMHT2014} & 1.0 $\pm$ 0.6 & 2.6 $\pm$ 0.6  \\
        \texttt{NNPDF3.1} & 1.1 $\pm$ 0.3 & 2.1 $\pm$ 0.4  \\
        \texttt{CTEQ6M} & N/A  & 2.8 $\pm$ 0.9 \\
        \hline
    \end{tabular}    
    \caption{Comparison of the shift of $M_W$ for different PDF sets using the $m_T$ observable. The central prediction used was \texttt{CT18NNLO} with a mass of 80,385 MeV. The uncertainties quoted are the PDF uncertainties for the given PDF set.}
    \label{tab:pdf}
\end{table}

In conclusion, two of the major criticisms leveled against the theory calculations involved in the \textsc{ResBos} program cannot explain the deviation from the SM that is reported by CDF. We found that the data-driven techniques used by the CDF experiment help to reduce the effects of higher order corrections. The estimated shift due to including these corrections is at most 10 MeV, and may reduce the disagreement from 7$\sigma$ to 6$\sigma$. The PDF uncertainty is found to be consistent with the numbers quoted by CDF.
While there remain main questions pertaining to the CDF result, we have addressed the most important questions related to the theory calculations used, and have found no major mass shift.

\begin{acknowledgments}
We thank John Cambell for help with the MCFM code. We thank Simone Amoroso, Joshua Bendavid, Maarten Boonekamp, Tao Han, Joey Huston, Ashutosh Kotwal, Pavel Nadolsky, Boris Tuchming, Mika Anton Vesterinen, Michael Wagman, and Feng Yuan for helpful discussions. We thank Stefan H\"oche and Pedro Machado for comments on the manuscript. This manuscript has been authored by Fermi Research Alliance, LLC under Contract No. DE-AC02-07CH11359 with the U.S. Department of Energy, Office of Science, Office of High Energy Physics. It is also supported in part by the U.S.~National Science Foundation
under Grant No.~PHY-2013791. C.-P.~Yuan is also grateful for the support from the Wu-Ki Tung endowed chair in particle physics.
\end{acknowledgments}

\appendix

\section{Collins-Soper-Sterman Formalism}\label{app:resum}

There are many different formalisms used to perform the resummation of the transverse momentum of color singlet final states, including the CSS formalism~\cite{Collins:1977iv,Collins:2011zzd}, the CFG formalism~\cite{Catani:2000vq}, resummation in direct $p_T$ space~\cite{Bizon:2017rah}, SCET formalisms~\cite{Becher:2012yn,Chen:2018pzu,Becher:2020ugp}, and TMD formalisms~\cite{Rogers:2015sqa,Angeles-Martinez:2015sea}. The \textsc{ResBos2} code implements both the CSS and CFG formalism (both of which are closely related to the TMD formalism).
The version of the \textsc{ResBos} code used by the CDF collaboration only implements the CSS formalism, the shift in the $W$ mass induced by different formalisms is an ongoing work. 

The CSS formalism was introduced in Ref.~\cite{Collins:1977iv}, and uses impact parameter space to formally resum the
logarithmic terms in the fixed order calculation. This involves solving the renormalization group equations (RGE) given by
\begin{align}\label{eq:rge}
\frac{d}{d\log\mu}K\left(b\mu,g_s\left(\mu\right)\right) &= -\gamma_K\left(g_s\left(\mu\right)\right)K\left(b\mu,g_s\left(\mu\right)\right), \\
\frac{d}{d\log\mu}G\left(b/\mu,g_s\left(\mu\right)\right) &= \gamma_K\left(g_s\left(\mu\right)\right)G\left(b/\mu,g_s\left(\mu\right)\right),
\end{align}
where $\gamma_K$ is the anomalous dimension given by
\begin{equation}
 \gamma_K = \Gamma_{\text{cusp}}\log(Q^2/\mu^2)+ \gamma_i
\end{equation}
The logarithms that are resummed are determined by the order of the cusp ($\Gamma_{\text{cusp}}$) and non-cusp ($\gamma_i$) annomalous dimensions. These result in the order of the $A$ and $B$ coefficients in the CSS formalism (see Eq.\eqref{eq:sud}). The values for $A^{(1)}$, $A^{(2)}$, $A^{(3)}$, $B^{(1)}$ and $B^{(2)}$ are given as
\begin{align}
    A^{(1)} &= C_F, \\
    A^{(2)} &= \frac{1}{2}C_F\left(\left(\frac{67}{18}-\frac{\pi^2}{6}\right)C_A-\frac{5}{9}N_f\right), \\
    A^{(3)} &= C_F\left(\frac{C_F N_f}{2}\left(\zeta_3-\frac{55}{48}\right)-\frac{N_f^2}{108}\right.\\ 
    &\left.+C_A^2\left(\frac{11\zeta_3}{24}+\frac{11\pi^4}{720}-\frac{67\pi^2}{216}+\frac{245}{96}\right) \right.\nonumber \\
    &\left.+C_A N_f\left(\frac{-7\zeta_3}{12}+\frac{5\pi^2}{108}-\frac{209}{432}\right)\right),\nonumber \\
    B^{(1)} &= -\frac{3}{2}C_F, \\
    B^{(2)} &= C_F^2\left(\frac{\pi^2}{4}-\frac{3}{16}-3\zeta_3\right)\\
    &+C_F C_A\left(\frac{11}{36}\pi^2-\frac{193}{48}+\frac{3}{2}\zeta_3\right)\nonumber \\
    &+C_F N_f\left(\frac{17}{24}-\frac{\pi^2}{18}\right), \nonumber
\end{align}
where $C_F=4/3$, $C_A=3$, and $N_f$ is the number of active quarks used in the calculation of the $\beta$ function. The values for $A^{(4)}$ are given in Refs.~\cite{Henn:2019swt,vonManteuffel:2020vjv}, and for $B^{(3)}$ in Ref.~\cite{Li:2016ctv}. The hard collinear coefficient at $\mathcal{O}(\alpha_s)$ is given as
\begin{align}
    C_{qq}^{(1)}(z) &= \frac{1}{2}C_F(1-z) + \delta(1-z)\frac{1}{4}C_F\left(\pi^2-8\right),\\
    C_{qg}^{(1)}(z) &= \frac{1}{2}z(1-z), \\
    C_{q\bar{q}}^{(1)}(z) &= C_{qq'}^{(1)}(z) = C_{q\bar{q}'}^{(1)}(z) = 0,
\end{align}
and the coefficient at $\mathcal{O}(\alpha_s^2)$ are given in Ref.~\cite{Catani:2012qa}.

The scale variations are induced by varying the parameters $C_2$, $C_1/b_0 = C_3/b_0 = \mu_F$, and $\mu_R$ within the calculation, where $b_0=2 e^{-\gamma_E}$ and $\gamma_E$ is the Euler constant. The parameters $C_{1,2,3}$ arise from solving the RGE, while $\mu_R$ and $\mu_F$ are the typical renormalization and factorization scales, in the unit of $\sqrt{Q^2+p_T^2}$. The scales are varied by factors of two around the central values, with the constraint that the ratio of any two scales is not greater than 2 or less than 0.5. This results in a total of 15 different scale variations. Additionally, to address the fact that the scale for $C_3$ appears in the PDFs, we choose to enforce that this scale remains in the perturbative region and only shift down to 0.7 of the central value, instead of by a factor of 2.

\subsection{Non-Perturbative Function}\label{app:nonpert}

In the resummation calculation, the lower limit of the integral in Eq.~\eqref{eq:sud} results in evaluating the strong coupling constant at a scale in which it becomes non-perturbative. To handle this issue, a prescription needs to be introduced to ensure the perturbative calculation remains perturbative. In this work, we use the $b^*$ prescription~\cite{Collins:1977iv}
\begin{equation}
    b^* = \frac{b}{\sqrt{1+\frac{b^2}{b_{\text{max}}^2}}}\, ,
\end{equation}
where $b_{\text{max}}$ is a parameter fixed to determine the maximum allowed $b$ value. To be consistent with the calculation used by CDF, we choose to take $b_{\text{max}}=0.5$ GeV${}^{-1}$. This converts Eq.~\eqref{eq:sud} into two contributions
\begin{align}
    S(b) &= S_{\text{NP}}S_{\text{Pert}}\,, \\
    S_{\text{Pert}}(b) &= \int^{C_2^2 Q^2}_{C_1^2/(b^*)^2} \frac{{\rm d}\bar{\mu}^2}{\bar{\mu}^2} \left[\ln\left(\frac{C_2^2 Q^2}{\bar{\mu}^2}\right) A\left(\bar{\mu}, C_1\right) \right. \nonumber\\
    &\left.+ B\left(\bar{\mu}, C_1, C_2\right)\right.\bigg]\, . \nonumber
\end{align}
The functional form of $S_{NP}$ is fit to data, and the parametrization used is inspired by the evolution of the Collins-Soper kernel.

CDF used the Brock-Landry-Nadolsky-Yuan (BLNY) non-perturbative functional form fit to global data within their analysis~\cite{Landry:1999an,Landry:2002ix}. The functional form is given as:
\begin{equation}
S_{\text{NP}} = \left[-g_1-g_2\ln\left(\frac{Q}{2Q_0}\right)-g_1g_3\ln\left(100x_1x_2\right)\right]b^2\, ,
\end{equation}
where $Q_0$ is fixed to 1.6 GeV.
The parameters $g_1$, $g_2$, and $g_3$ are
constrained from the combined fit to the low transverse momentum
distributions of  Drell-Yan lepton pair production
with $4\ \text{GeV} < Q < 12\ \text{GeV}$ in fixed
target experiments and $W$ and $Z$ production ($Q\sim 90\ \text{GeV}$)
at the Tevatron. The best fit gives the values: $g_1=0.21$, $g_2=0.68$, and $g_3=-0.60$~\cite{Landry:1999an,Landry:2002ix}. In the BLNY work, there was no consideration about the relationship between 
the non-perturbative parameters $g_{1,2,3}$ and the choice of the perturbative scales  $C_1$ and $C_3$. In principle, varying these scales changes the relative amount of contributions from  non-perturbative long-distance physics and the perturbative short-distance physics. Therefore, the non-perturbative function should depend on the choice of $C_1$ and $C_3$. The investigation of this effect is left to a future work.

An investigation into the impact of flavor dependence of the non-perturbative function and the choice of functional form on the extracted $W$ mass is left to a future work. Some preliminary estimates for the flavor dependence effects for the ATLAS measurement were studied in Ref.~\cite{Bacchetta:2018lna}. Additionally, lattice QCD calculations of TMDs, the non-perturbative function, and flavor dependence are an active area of research~\cite{Musch:2010ka,Musch:2011er,Engelhardt:2015xja,Yoon:2017qzo,Shanahan:2019zcq,Shanahan:2020zxr,LatticeParton:2020uhz,Schlemmer:2021aij,Li:2021wvl,Shanahan:2021tst,LPC:2022ibr}. Detailed studies on the impact of these results on the extraction of the $W$ mass are needed.

\section{Angular Coefficients}\label{app:angular}

\begin{figure}
    \centering
    \includegraphics[width=0.47\textwidth]{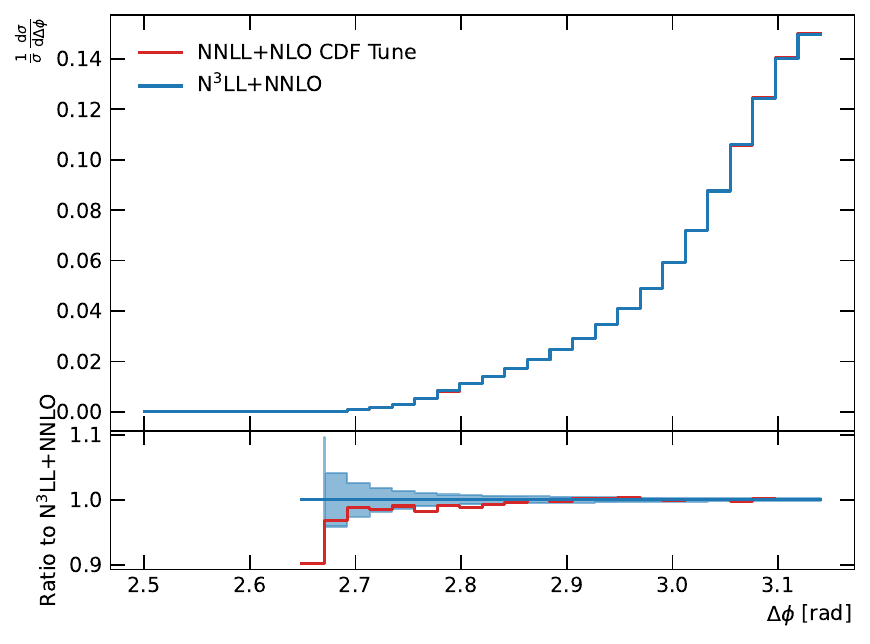}
    \caption{Comparison of the generated pseudodata for $\Delta\phi$ using the N${}^3$LL+NNLO calculation compared to the CDF tuned prediction at NNLL+NLO. The blue band represents the statistical uncertainty associated with the CDF measurement.}
    \label{fig:delta_phi}
\end{figure}

As previously discussed, the \textsc{ResBos2} code uses MCFM~\cite{Campbell:2015qma} to correct the angular coefficients to be accurate at NNLO. The impact of these coefficients at the Tevatron on the $W$ mass extraction are expected to be small. The most important distribution to consider when measuring the $W$ mass at the Tevatron is the difference in azimuthal angle between the lepton and the missing momentum in the lab frame used in calculating $m_T$ ($\Delta\phi$), see Eq.~\eqref{eq:mt}. Figure~\ref{fig:delta_phi} shows the comparison between the result for N${}^{3}$LL+NNLO accuracy compared to NNLL+NLO accuracy. The uncertainty is the statistical uncertainty quoted by CDF. We find that this distribution is stable to the order of the calculation, and thus changing the order would not have a noticeable impact on the $W$ mass measurement.
A comparison of the angular coefficients to the results from the LHC is left to a future work.

\section{Width Effects}\label{app:width}

\begin{figure}[htpb]
	\centering
	\includegraphics[width=0.47\textwidth]{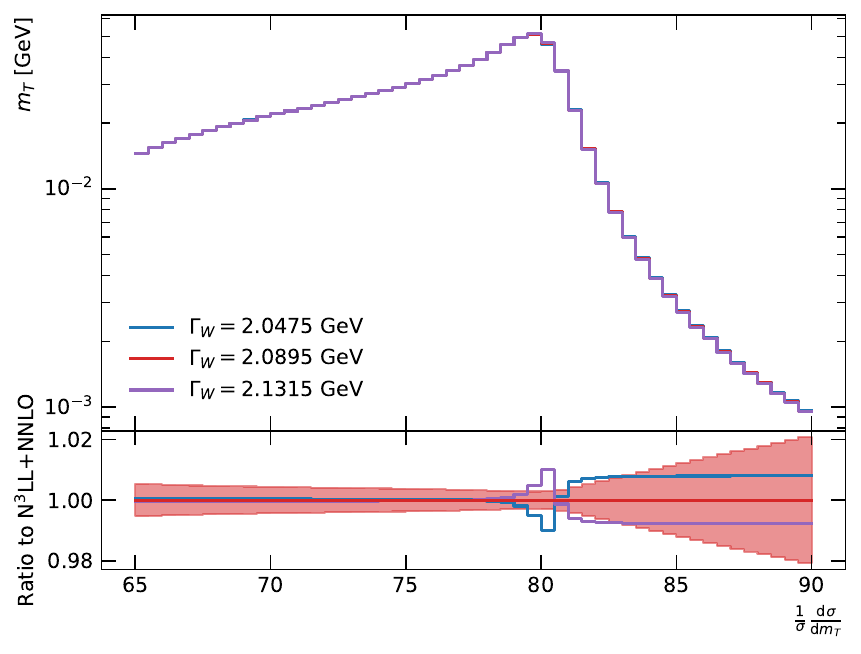}
	\caption{Comparison of the $m_T$ distribution for various different choices of $\Gamma_W$. The width used by CDF was 2.0895 GeV (red curve), and the blue and purple curve represent the shift in the width up and down by one standard deviation of the uncertainty quoted by the PDG~\cite{ParticleDataGroup:2020ssz}.}
	\label{fig:width}
\end{figure}

In the extraction of the $W$ mass, the CDF experiment kept fixed the width of the $W$ boson to 2.0895 GeV. We follow the approach taken by CDF and write the propagator of the $W$ boson as a Breit-Wigner shape with an energy-dependent width. The couplings of gauge bosons to fermions are defined in the $G_\mu$ scheme.
To estimate the impact of varying the width on the CDF result, we varied the width by 0.042 GeV based on the uncertainty from the global width measurement from the PDG~\cite{ParticleDataGroup:2020ssz}. Additionally, a fourth variation was used in which the width was fixed to the Standard Model prediction for the width at NLO, in which the width is proportional to $M_W^3$. The experimental observable most sensitive to the width is $m_T$, and thus we only preformed the extraction for this observable. The effect of the width on $m_T$ can be found in Fig.~\ref{fig:width}, where the red uncertainty band gives the statistical uncertainty from CDF. It is clear that the effect of the width is important at high $m_T$ (\textit{i.e.} $m_T > 80$ GeV). Table~\ref{tab:width} shows the extracted mass shift for the different mass scenarios described above.

\begin{table}[htpb]
	\centering
	\begin{tabular}{|c|c|}
		\hline
		Width & Mass Shift [MeV] \\
		\hline
		2.0475 GeV & 2.0 $\pm$ 0.5 \\
		2.1315 GeV & 0.3 $\pm$ 0.5 \\
		NLO & 1.2 $\pm$ 0.5 \\
		\hline
	\end{tabular}
	\caption{The shift in $M_W$ due to changing the width. The width is varied by the uncertainty from the PDG~\cite{ParticleDataGroup:2020ssz}, with the central value set to 2.0895 GeV used by the CDF collaboration~\cite{CDF:2022hxs}. Additionally, the Standard Model prediction for the width at NLO is considered.}
	\label{tab:width}
\end{table}

\section{The ratio for the normalized $p_T(W)$ to the normalized $p_T(Z)$}
\label{app:ratio_pT}

\begin{figure}[htbp]
	\includegraphics[width=0.47\textwidth]{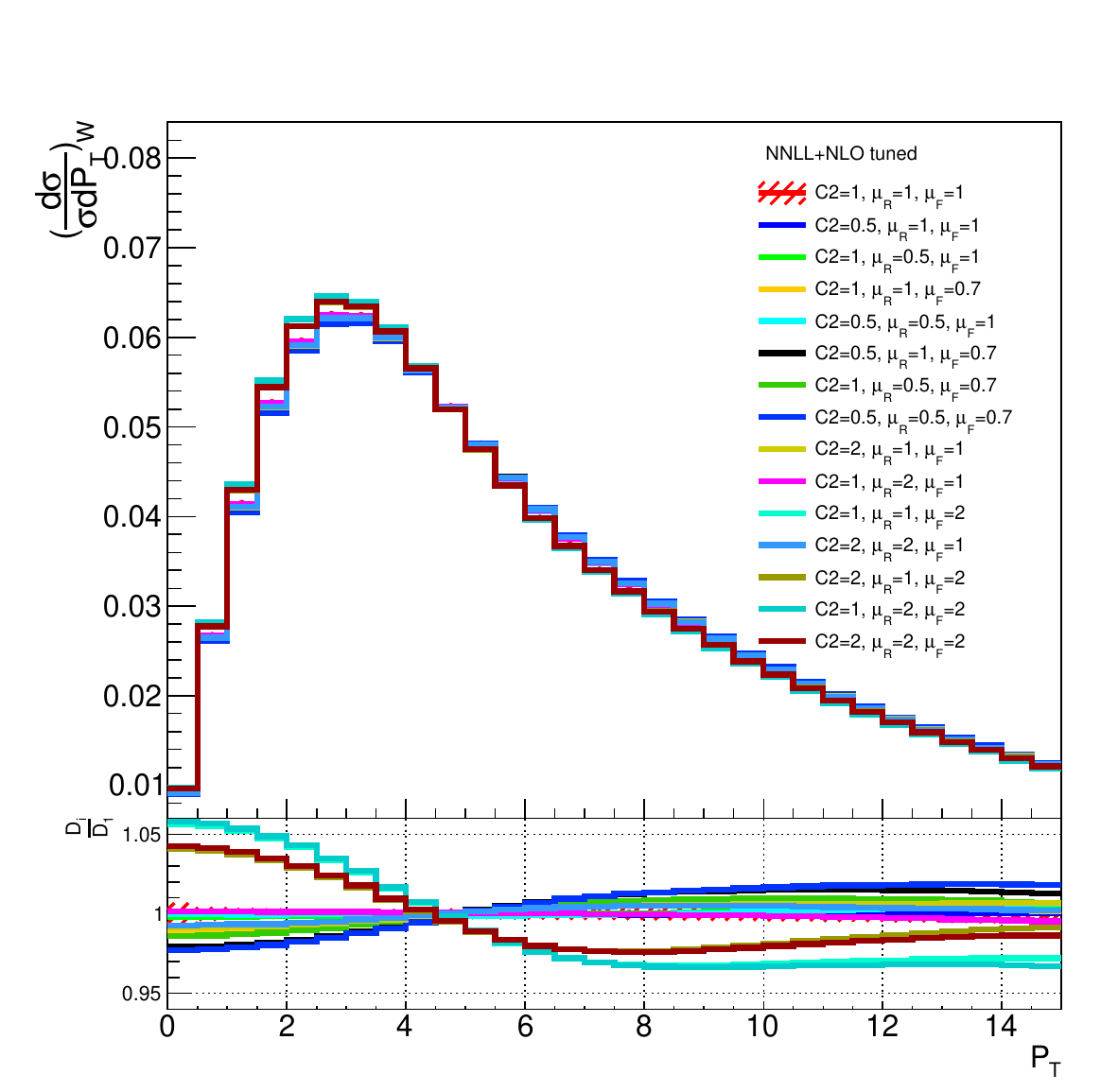}
	\caption{Normalized $P_T(W)$ distributions due to QCD scale variation.}
	\label{fig:pTW_15scale}
\end{figure}

A new feature in the recent CDF measurement was to use the ratio of the normalized $p_T(W)$ distribution to the normalized $p_T(Z)$ distribution to further constrain their systematic uncertainties. To estimate the theory uncertainty in this ratio, the CDF experiment used the DYQT code~\cite{Bozzi:2008bb,Bozzi:2010xn} instead of the ResBos code.
Here, we follow the procedure carried out by CDF. In this study, the scales are treated as fully correlated between $p_T(W)$  and $p_T(Z)$ 
in the ``tuned'' ResBos NNLL+NLO predictions.
Figure~\ref{fig:pTW_15scale} shows in the upper panel the normalized $p_T(W)$ distribution, $D_i=({d \sigma \over \sigma d p_T})_W(i)$, for each of the 15 scale choices, and in the lower panel the ratio of $D_i/D_1$, where $D_1$ corresponds to the canonical scale choice. By applying the envelope method~\cite{CDF:2022hxs}, CDF further constrained the allowed QCD scale variation in the normalized $p_T(W)$ distribution by fitting to their measurement of the normalized $p_T(W)$ distribution with $\Delta \chi^2=1$. 
Given the allowed range of the normalized $p_T(W)$ distribution, one could extract the fitted value of $M_W$ from the corresponding $m_T$, $p_T(\ell)$ and $p_T(\nu)$ distributions, respectively. This result is shown in Table~\ref{tab:mw_envelope}.
We have checked that a similar conclusion also holds when using the N${}^3$LL+NNLO predictions.

\begin{table*}[ht]
	\centering
	\scalebox{0.9}{
		\begin{tabular}{|c|c|c|c|c|c|c|}
			\hline
			& \multicolumn{6}{c|}{Mass Shift [MeV]} \\
			\hline
			& \multicolumn{2}{c|}{$m_T$} & \multicolumn{2}{c|}{$p_T(\ell)$} & \multicolumn{2}{c|}{$p_T(\nu)$} \\
			\hline
			Scale & \textsc{ResBos2} & +Detector Effect+FSR & \textsc{ResBos2} & +Detector Effect+FSR & \textsc{ResBos2} & +Detector Effect+FSR \\
			\hline
			Upper & 1.2 $\pm$ 0.5 & 0.8 $\pm$ 1.8 $\pm$ 1.1 & 3.1 $\pm$ 2.1 & -6.5 $\pm$ 2.7 $\pm$ 1.3 & 1.4 $\pm$ 2.1 & -4.9 $\pm$ 3.4 $\pm$ 2.0 \\
			\hline
			Lower & 1.2 $\pm$ 0.5 & -0.7 $\pm$ 1.8 $\pm$ 01. & 1.8 $\pm$ 2.1 & 9.4 $\pm$ 2.6 $\pm$ 1.2 & 0.0 $\pm$ 2.1 & 4.8 $\pm$ 3.4 $\pm$ 1.9 \\
			\hline
		\end{tabular}
	}
	\caption{The shift in $M_W$ due to QCD scale variation in the ratio of the normalized $p_T(W)$ distribution to the normalized $p_T(Z)$ distribution, following the envelope method carried out by CDF.}
	\label{tab:mw_envelope}
\end{table*}

\section{Detector Smearing and Final State Radiation Effects}\label{app:detector}

\begin{figure}[htpb]
    \centering
    \includegraphics[width=0.47\textwidth]{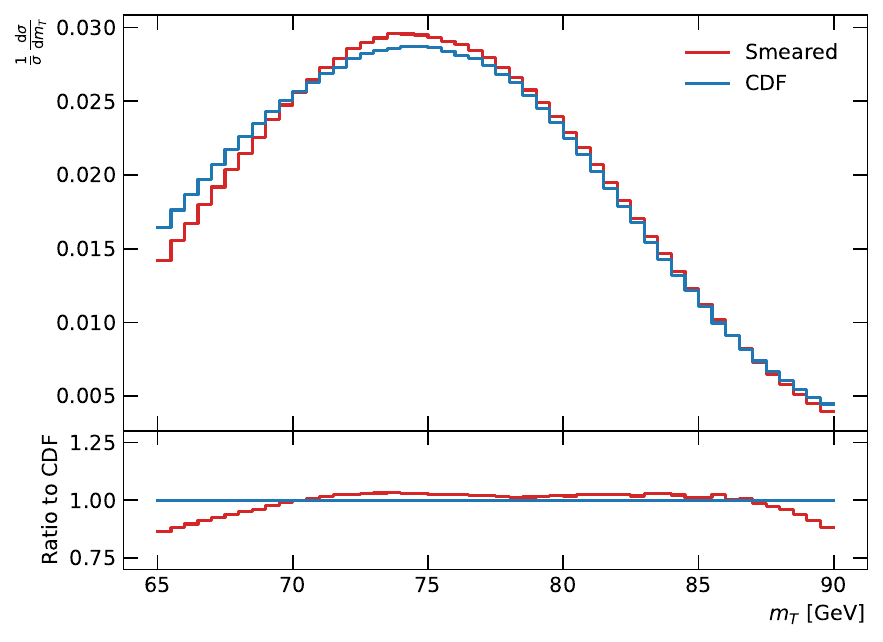}
    \caption{Comparison of the smeared $m_T$ distribution to the CDF data. The red curve is the result of the smearing, and the blue curve is the extracted CDF data in the electron channel.}
    \label{fig:detector_smear}
\end{figure}

The \textsc{ResBos} code does not contain final state QED radiation (FSR) nor detector effects in the calculation. These effects smear out the observables used to measure the $W$ mass, especially $m_T$. To investigate the impact of these effects, we parameterize the smearing effect and fit it to the data observed by CDF. In particular, we chose to study the electron channel since the impact of FSR and the background are both smaller than the muon channel. The comparison of the tuned smearing to the CDF data can be seen in Fig.~\ref{fig:detector_smear}. We used a three parameter fit to determine the width of the Gaussian based on the energy of the particle given by
$
\frac{\sigma}{E} = a \oplus \frac{b}{\sqrt{E}} \oplus \frac{c}{E},
$
where $a$, $b$, and $c$ are the parameters of the fit, and the terms are added in quadrature to obtain the width. Both the pseudodata and the mass templates are smeared using the same functional form, and the results are given in Tab.~\ref{tab:HOresults}. This form will not capture the detection efficiency nor all of the FSR effects, but can reproduce the general shape of the $m_T$ distribution shown in Fig.~\ref{fig:detector_smear}. We leave a detailed study of FSR to a future work, and require additional information from the CDF collaboration to accurately model the detector acceptance.

To estimate the impact of the approximation of the detector response on the extraction of $M_W$, we preform an additional check in which a simple Gaussian with a width of 5\% for electrons and 11\% for neutrinos is used to smear the results. The comparison between the two approaches are given in Tab.~\ref{tab:detector}. We find that the choice of smearing does not have an impact on the extracted result of $M_W$, so long as the smearing of the data and the templates are identical. The accuracy of the model used by CDF to smear the theory templates is beyond the scope of this work, but the inaccuracy of our model compared to that used by CDF does not change the conclusions drawn within this work.

\begin{table}[htpb]
    \centering
    \begin{tabular}{|c|c|c|}
        \hline
        & \multicolumn{2}{c|}{Mass Shift [MeV]} \\
        \hline
        Observable & Smearing 1 & Smearing 2 \\
        \hline
        $m_T$ & 0.2 $\pm$ 1.8 $\pm$ 1.0 & 1.0 $\pm$ 2.1 $\pm$ 1.3 \\
        $p_T(\ell)$ & 4.3 $\pm$ 2.7 $\pm$ 1.3 & 4.5 $\pm$ 2.6 $\pm$ 1.4  \\
        $p_T(\nu)$ & 3.0 $\pm$ 3.4 $\pm$ 2.2 & 3.8 $\pm$ 4 $\pm$ 2.7\\
        \hline
    \end{tabular}
    \caption{Summary of the shift in $M_W$ due to two different smearing methods. The first uncertainty denotes the statistical uncertainty, and the second uncertainty results from an approximate model simulating the detector effect and FSR, calculated from generating 100 different smearings on the data. ``Smearing 1'' refers to the fit result to the CDF data, and ``Smearing 2'' refers to the crude Gaussian smearing.}
    \label{tab:detector}
\end{table}

\section{PDF-induced Correlations}
\label{app:pdf_correlations}

\begin{figure}
    \centering
    \includegraphics[width=0.47\textwidth, clip, trim=10mm 25mm 0mm 15mm]{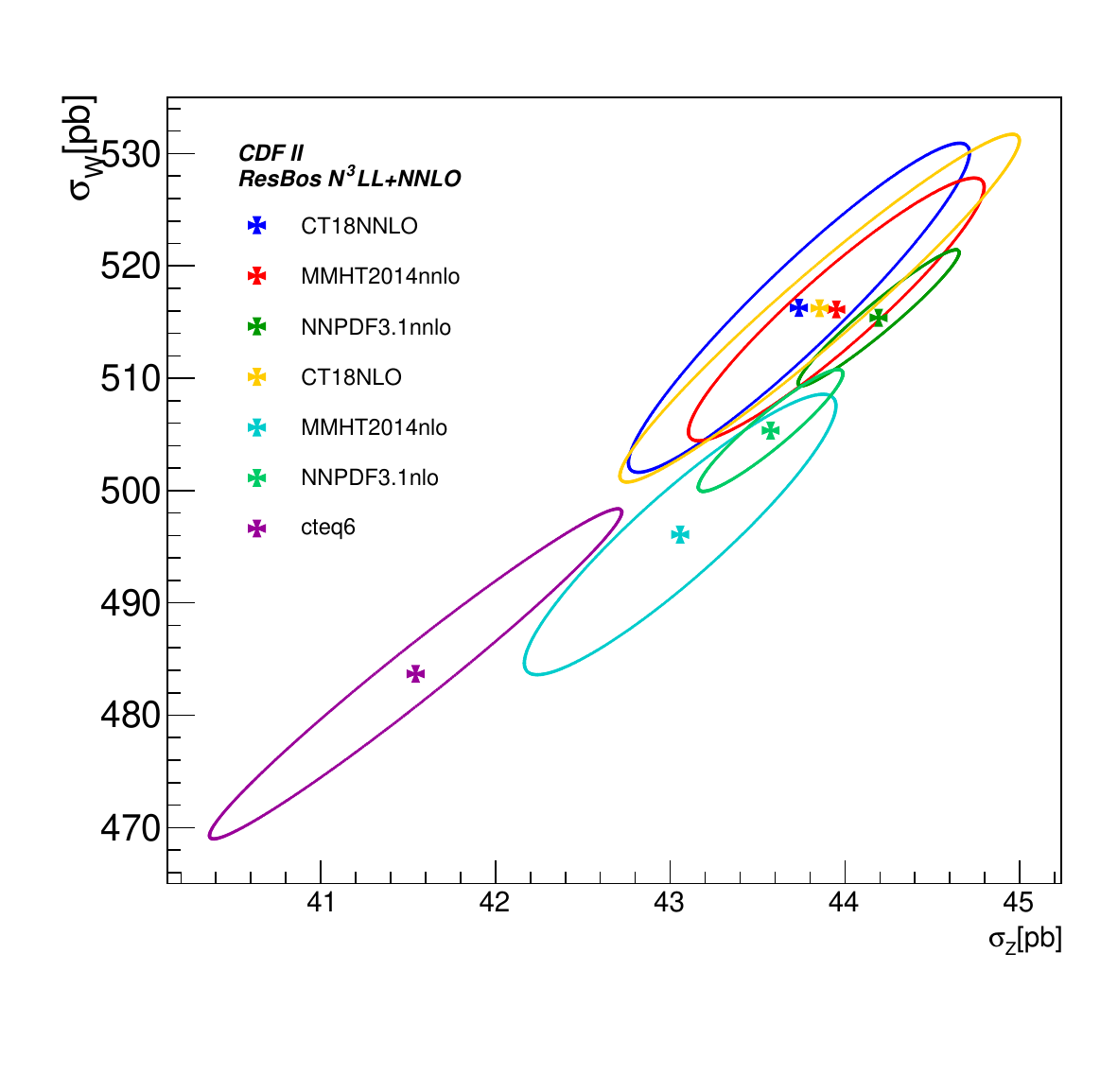}
    \caption{PDF-induced correlation ellipses, at the 68\% confidence level (C.L.), between the fiducial cross sections of $W$ and $Z$ boson  production at the Tevatron Run II.}
    \label{fig:Corr_ellipse}
\end{figure}

\begin{figure}
    \centering
    \includegraphics[width=0.47\textwidth, clip, trim=10mm 25mm 0mm 15mm]{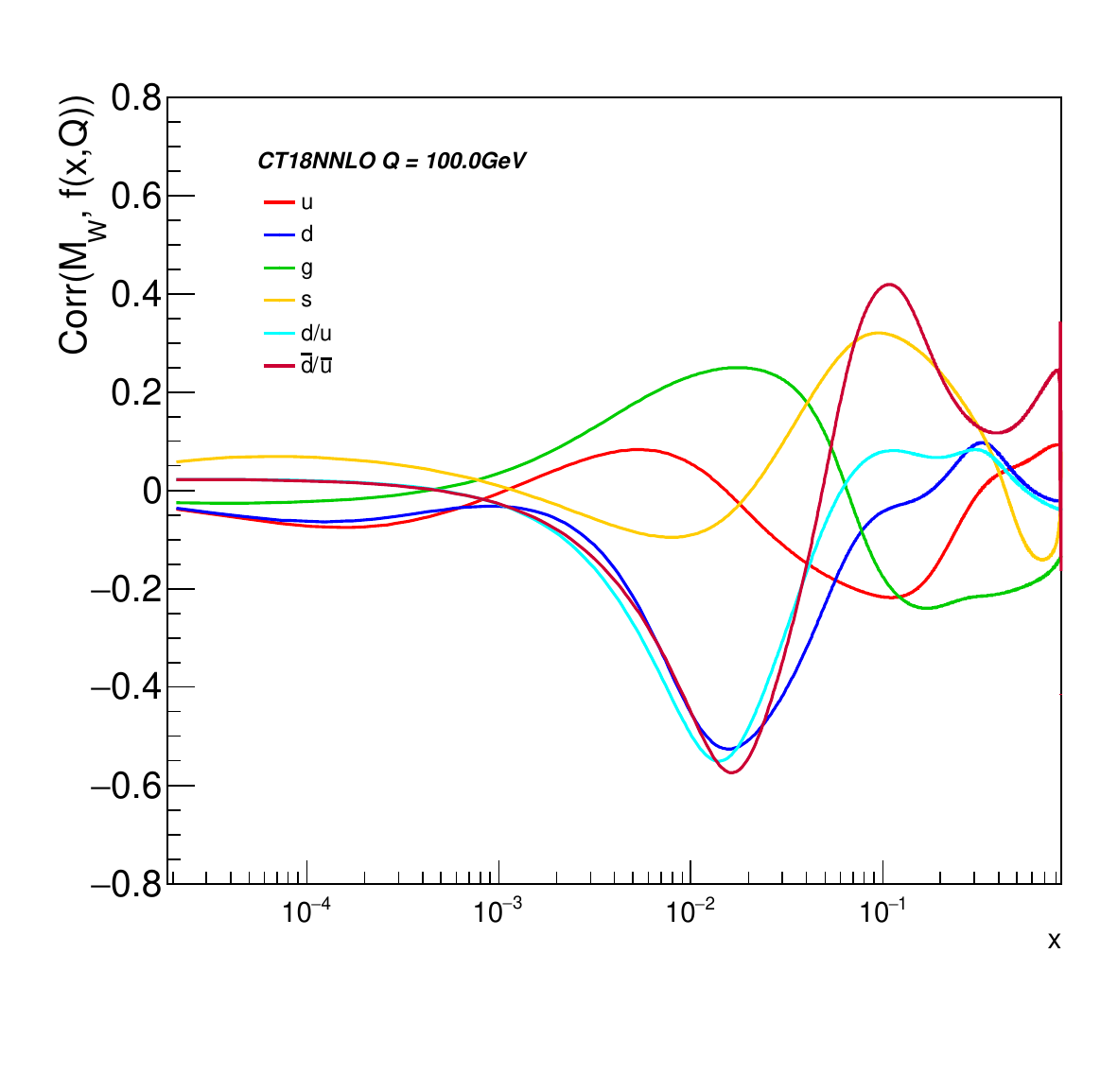}
    \caption{PDF-induced correlation cosine between the extracted $W$ boson mass (from $m_T$ distribution) and the CT18 NNLO PDFs at the specified $x$ value with $Q=100$ GeV. } 
   \label{fig:CorrCosine}
\end{figure}

In Table~\ref{tab:pdf}, we compared the shift of $M_W$ for different PDF sets using the $m_T$ observable. Similar comparisons using the $p_T(\ell)$ and $p_T(\nu)$ observables are listed in Table~\ref{tab:pdf_pt}. Again, the central prediction used was \texttt{CT18NNLO} with a mass of 80,385 MeV. The uncertainties quoted are the PDF uncertainties for the given PDF set. 
As noted in the main text, it is unclear to us how to appropriately propagate the uncertainties from individual observables to the final mass extraction done by CDF.

\begin{table*}[htbp]
    \centering
    \begin{tabular}{|c|c|c|c|c|c|c|}
        \hline
         &\multicolumn{2}{c|}{$m_T$} & \multicolumn{2}{c|}{$p_T(\ell)$} & \multicolumn{2}{c|}{$p_T(\nu)$}\\
        \hline
        PDF Set & NNLO & NLO  & NNLO & NLO & NNLO & NLO \\
        \hline
        \texttt{CT18} & 0.0 $\pm$ 1.3 & 1.8 $\pm$ 1.2 & 0.0 $\pm$ 15.9 & 2.0 $\pm$ 14.3 & 0.0 $\pm$ 15.5 & 2.9 $\pm$ 14.2 \\
        \texttt{MMHT2014} & 1.0 $\pm$ 0.6 & 2.6 $\pm$ 0.6 & 6.2 $\pm$ 7.8 & 36.7 $\pm$ 7.0 & 3.9 $\pm$ 7.5 & 36.0 $\pm$ 6.7 \\
        \texttt{NNPDF3.1} & 1.1 $\pm$ 0.3 & 2.1 $\pm$ 0.4 & 2.1 $\pm$ 3.8 & 13.5 $\pm$ 4.9 & 5.4 $\pm$ 3.7 & 10.0 $\pm$ 4.9 \\
        \texttt{CTEQ6M} & N/A  & 2.8 $\pm$ 0.9 & N/A & 19.0 $\pm$ 10.4 & N/A & 20.9 $\pm$ 10.2 \\
        \hline
    \end{tabular}  
	\caption{Comparison of the shift of $M_W$ for different PDF sets using the $m_T$, $p_T(\ell)$ and $p_T(\nu)$ observables, respectively. The central prediction used was \texttt{CT18NNLO} with a mass of 80,385 MeV. The uncertainties quoted are the PDF uncertainties for the given PDF set.
	}
	\label{tab:pdf_pt}
\end{table*}

Below, we briefly summarize a few PDF-induced correlations, predicted at N$^3$LL+NNLO, relevant to the CDF analysis. 

Fig.~\ref{fig:Corr_ellipse} shows the PDF-induced correlation ellipses between the fiducial cross sections of $W$ and $Z$ boson productions ($\sigma_W$ vs. $\sigma_Z$) at the Tevatron Run II for various PDF sets. Here, the kinematic cuts, as discussed in the main text, have been imposed. Fig.~\ref{fig:CorrCosine} displays the PDF-induced correlation of $M_W$ (extracted from the $m_T$ distribution) and CT18 NNLO error PDFs, for various flavors as a function of $x$ at $Q=100$ GeV. It shows that at the typical value of $x$, for the inclusive production of $W$ boson at the Tevatron, around $M_W/\sqrt{S} \simeq 80/1960 \simeq 0.04$, the PDF-induced error in $M_W$ is mainly correlated to that of the PDF-ratios ${\bar d} / {\bar u}$, $d/u$ and $d$-PDFs. 

\begin{figure}
    \centering
	\includegraphics[width=0.47\textwidth, clip, trim=10mm 25mm 0mm 10mm]{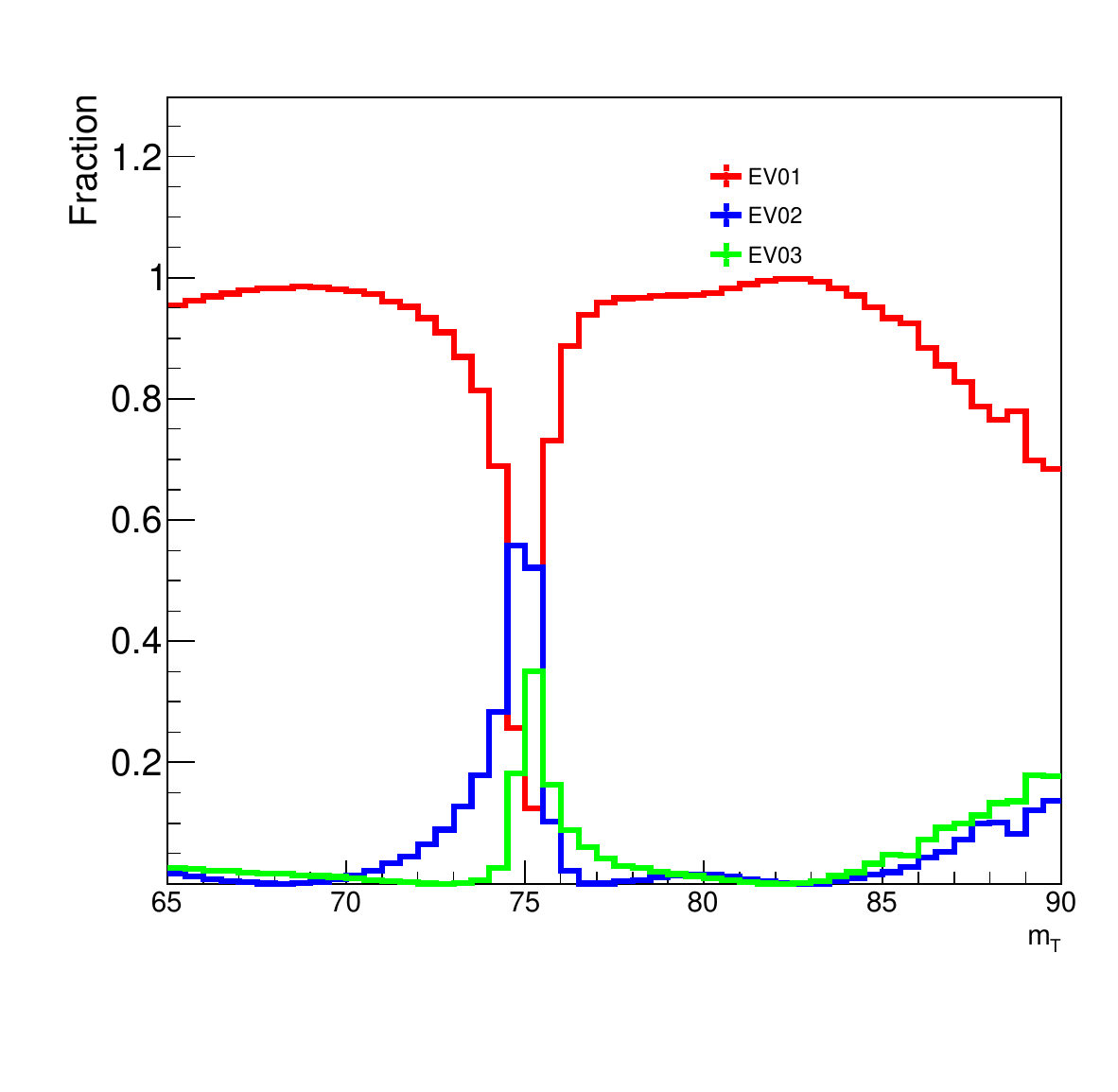}
	\caption{Fractional contribution of the three leading optimized eigenvector PDFs (EV01, EV02 and EV03) to the variance of the $m_T$ distribution, normalized to each bin, obtained from the ePump-optimization analysis.    
}
	\label{fig:frac_ev}
\end{figure}

\begin{figure}
    \centering
    \includegraphics[width=0.22\textwidth]{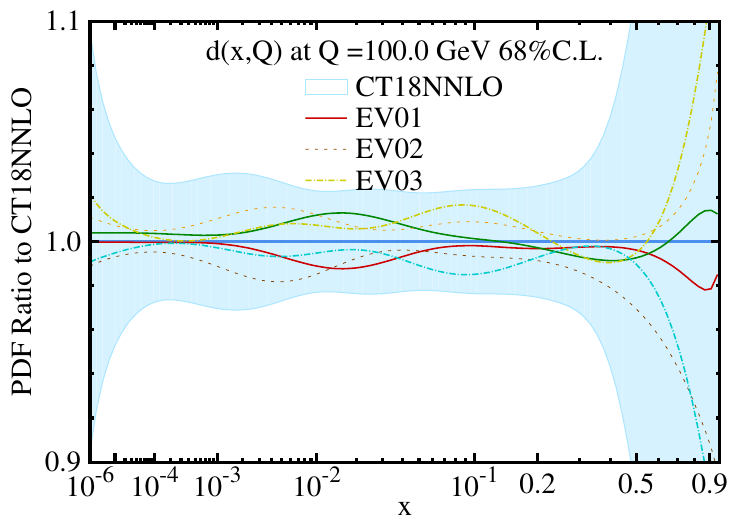}    
    \includegraphics[width=0.22\textwidth]{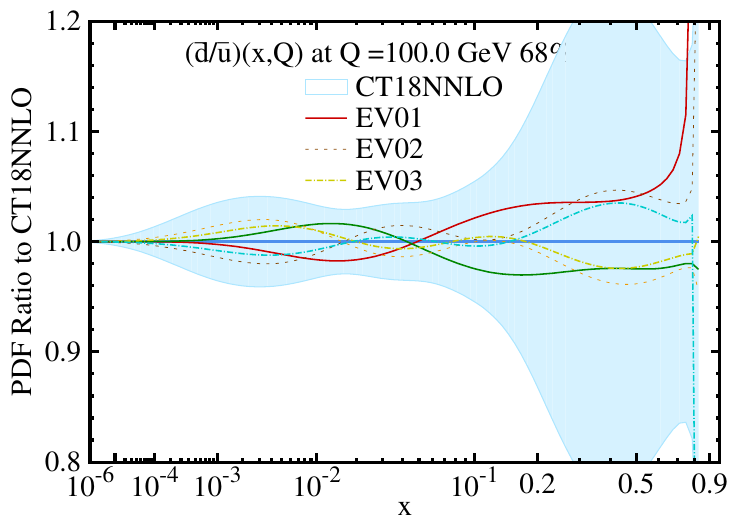} \\     \includegraphics[width=0.22\textwidth]{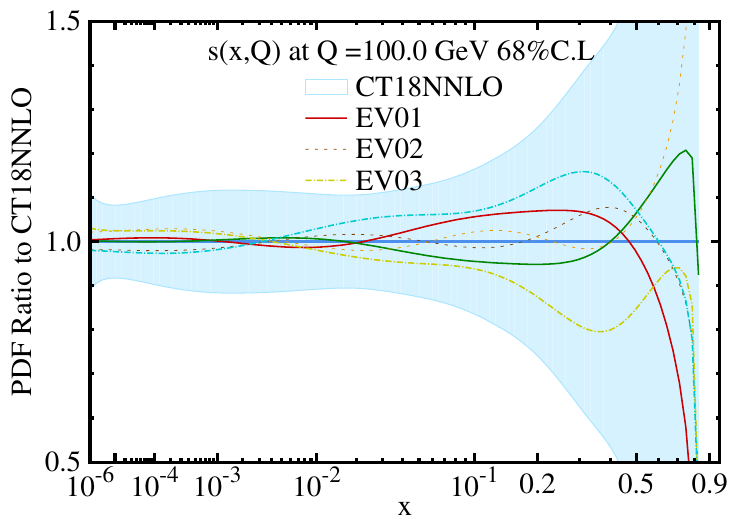}
    \includegraphics[width=0.22\textwidth]{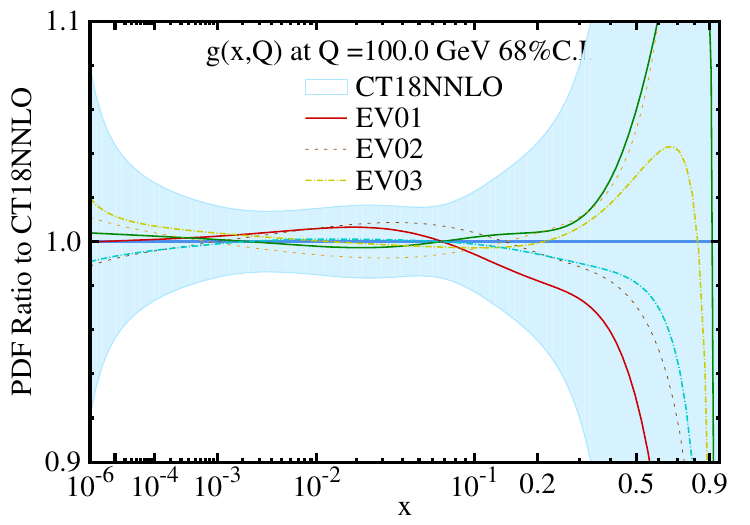}
    \caption{Ratios of the top three pairs of eigenvector PDFs and the original CT18 NNLO error PDFs, at $Q=100$ GeV, to the CT18 NNLO central value of $d$,  $\bar d/\bar u$, $s$ and $g$  PDFs. These eigenvector PDFs were obtained after applying the ePump-optimization to the original CT18 NNLO PDFs with respect to the $m_T$ distribution.}
    \label{fig:PDFFlavor}
\end{figure}

As discussed in Refs.~\cite{Schmidt:2018hvu,Hou:2019gfw}, one can easily find the first few leading eigenvector (EV) sets of error PDFs relevant to a particular experimental observable, such as the $m_T$ distribution, by applying the ePump-optimization procedure. This application of  \textsc{ePump} (error PDF Updating Method
Package) is based on ideas similar to that used in the data set diagonalization
method developed by Pumplin~\cite{Pumplin:2009nm}. It takes a set of Hessian error PDFs and constructs an equivalent set
of error PDFs that exactly reproduces the Hessian symmetric PDF uncertainties, but in addition each new eigenvector pair has an eigenvalue that quantitatively describes its contribution to the PDF uncertainty of a given data set or sets. 
The new eigenvectors can be considered as projecting the original error PDFs to the given data set, and be optimized or re-ordered so that it is easy to choose a reduced set that covers the PDF uncertainty for the input data set to any desired accuracy~\cite{Schmidt:2018hvu,Hou:2019gfw}.
The result is shown in Figs.~\ref{fig:frac_ev} and \ref{fig:PDFFlavor}. The eigenvalues of the three leading EV sets, after applying the ePump-optimization, are $44.5$, $3.0$, $2.4$, respectively. The combination of those top three optimized error PDFs contributes up to $99.6\%$ in the total PDF variance of the 50 given data points, {\it i.e.,} with 50 bins in $m_T$ distribution. This ePump-optimization allows us to conveniently use these three leading new eigenvectors (with a total of six error sets), in contrast to applying the full 58 error sets of the CT18 NNLO PDFs, to study the PDF-induced uncertainty of the  $m_T$ observable. Among them, the leading set EV01 dominates the Jacobian region, for $m_T$ around $M_W$. Hence, one could use those three leading pairs of EV sets to perform Monte Carlo study, such as studying the detector effect on the determination of $M_W$. The contributions provided by those three pairs of eigenvector PDFs to the PDF-induced uncertainty of $m_T$ distribution, for various parton flavors and $x$-ranges, are depicted in Fig.~\ref{fig:PDFFlavor}. The first eigenvector pair (EV01) gives the largest PDF contribution to the $m_T$ uncertainty, dominates the $d$ and ${\bar d} / {\bar u}$  uncertainties in the $x$ region of the $W$ boson production at the Tevatron. In the same figure, we also show the other two noticeable contributions of those three leading EV sets, which are found in the strangeness and gluon PDFs. 

\begin{figure}
    \centering
    \includegraphics[width=0.47\textwidth, clip, trim=10mm 25mm 0mm 15mm]{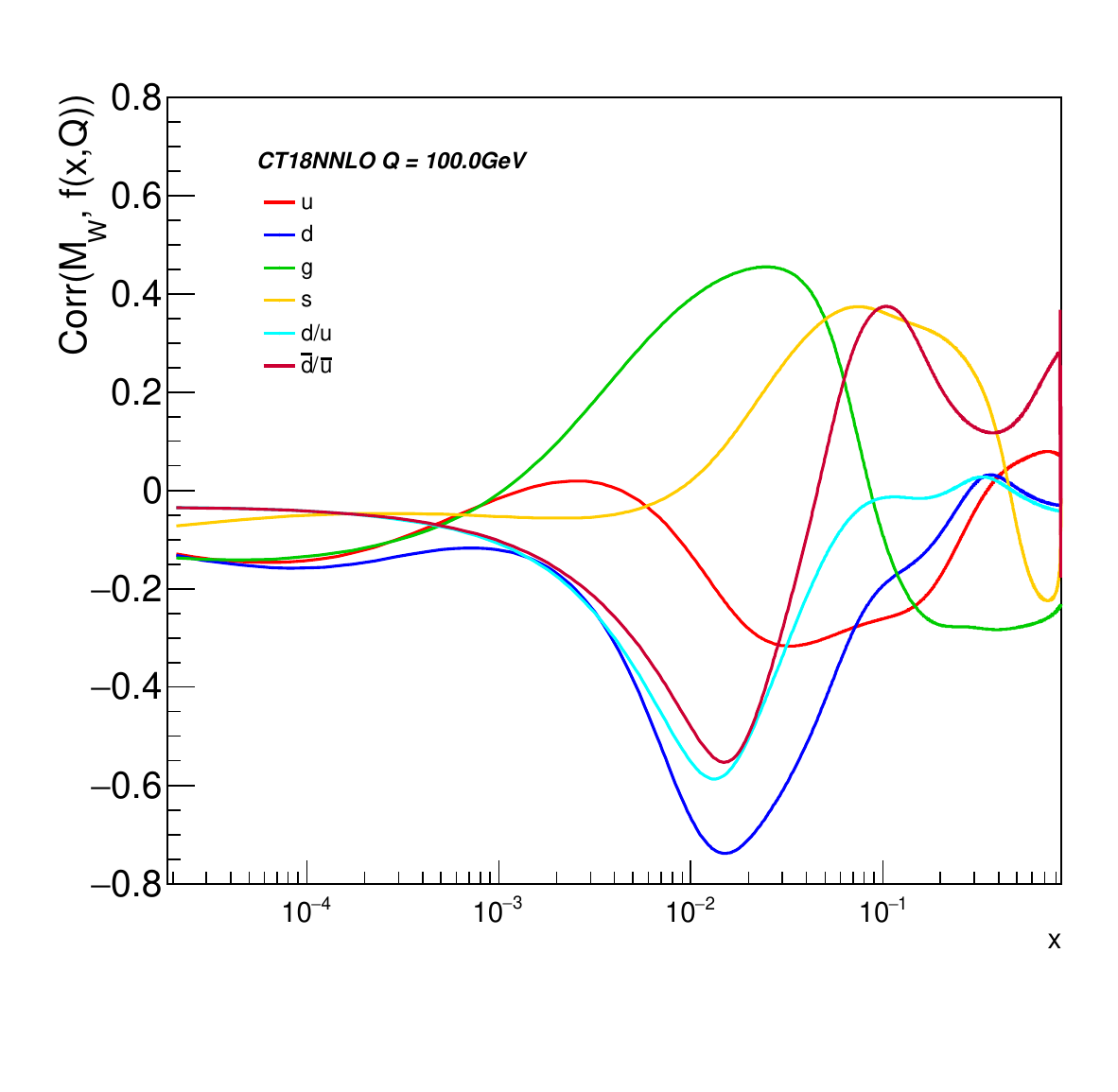}
    \caption{Similar to Fig.~\ref{fig:CorrCosine}, but from $p_T(\ell)$ distribution.}
   \label{fig:CorrCosine_Lep}
\end{figure}

For completeness, we also show in Fig.~\ref{fig:CorrCosine_Lep} the correlation cosine plot similar to Fig.~\ref{fig:CorrCosine}, but for the $W$ boson mass extracted from $p_T(\ell)$ distribution. (The same conclusion also holds for using the normalized $p_T(\nu)$ distribution.)
We note that the most relevant PDF flavors in these two cases are $d$ and $g$ PDFs. Due to the larger uncertainty in $g$-PDF, the PDF-induced uncertainty on the extracted $W$ boson mass is larger than the one extracted from the $m_T$ distribution, as shown in Table~\ref{tab:pdf_pt}.

\bibliography{biblio}
\end{document}